\documentclass[final]{IEEEtran}

\usepackage{widetext}
\usepackage{amsfonts}
\usepackage{graphicx}
\usepackage{amsmath}

\usepackage[noadjust]{cite}
\usepackage{subfig}

\begin{document}

\title{Diamond Configuration for Non-reciprocal Transmission}

\author{Sina~Khorasani%, ̃\IEEEmembership{Senior Member, ̃IEEE}
	\thanks{S. Khorasani is with \'{E}cole Polytechnique F\'{e}d\'{e}rale de Lausanne (EPFL), CH-1015, Lausanne, Switzerland. e-mail: sina.khorasani@epfl.ch.}
	\thanks{The author is on leave from School of Electrical Engineering, Sharif University of Technology, Tehran, Iran. e-mail: khorasani@sina.sharif.edu.}
	\thanks{This work has been supported in part by the Research Deputy of Sharif University of Technology as well as Laboratory of Photonics and Quantum Measurements at EPFL.}
	\thanks{Manuscript received December 22, 2016, revised May 25, 2017.}
}
	
%\markboth{IEEE Journal of Quantum Electronics JQE-134539-2016.R1}{{Khorasani}: Non-reciprocal transmission}

\maketitle

\begin{abstract}
A system scheme is presented which allows non-reciprocal wave transmission or directional amplification of electromagnetic signals, using a boxed four-node method. Edges represent strong hopping interactions and diagonals stand for weak parametric interactions. Using careful optimization of values for design parameters, we are able to obtain non-reciprocity in excess of 12dB and 130dB for intrinsic and extrinsic configurations at identical input/output frequencies. For the directional amplification, an isolation as high as 40dB is demonstrated with forward/backward gains of $\pm$20dB. Cascading two such systems potentially can offer high isolations at high gains.
\end{abstract}

\begin{IEEEkeywords}
	Quantum Optics, Nonreciprocity, Langevin Equations.
\end{IEEEkeywords}

\section{Introduction}

\IEEEPARstart{S}{ince} the advent of quantum optomechanics \cite{0,00} many exciting unprecedented applications have been demonstrated. Non-reciprocal optomechanics is being considered one of the most interesting effects, which is being actively investigated and pursued \cite{1,2,3,3a,4,5,6,6a,6b,7,7a,7b,7c,7d,7e,7f,7g,7h}. The importance of non-reciprocal devices is primarily in demonstration of physics of symmetry-breaking as well as laying grounds for directional amplification of quantum information. The field of non-reciprocal transmission of signals is also being explored in other related fields to wave propagation such as micro-ring resonators \cite{8}, two-dimensional electronics \cite{9}, acoustics \cite{10,11,12,13,14}, optics \cite{15,16,16a,16b,16c,17}, and superconductive circuits \cite{17a,17b,17c,17d,17e,17f,17g,17h,17i,17j,17jt}.

While fundamental laws of the nature mostly preserve parity-time symmetry \cite{8,17k,17l}, this does not disallow one to design non-Hermitian system scattering matrices, in such a way that signal transmission along opposite directions are inequal. This concept is a bit far from daily intuition and requires careful system design and engineering. It is known that non-reciprocal behavior needs a minimum of three interacting modes \cite{2,3}. However, such a minimal system cannot be non-reciprocal if two of interacting modes have the same frequency. In that sense, the resulting non-reciprocal behavior would be actually a uni-directional frequency conversion. That is why a majority of non-reciprocal systems operate through either up-conversion or down-conversion of frequency in a uni-directional way. The goal of this presented design is to keep identical input/output frequencies, while retaining high non-reciprocity and/or directional amplification.

\section{System Description}

The system under consideration consists of four interacting modes $\omega_{n}$ with $n=1,2,3,4$ at four input/output ports. We simplify the design by operation at two identical frequencies, as $\omega=\omega_1=\omega_3$ and $\Omega=\omega_2=\omega_4$. This requirement is imposed by the fact that the incoming signal at the input and outgoing signal at the output ports must have identical frequencies. Hence, ports 1 and 3 may serve respectively as the input and output, while ports 2 and 4 may be used for pumping. We consider these four modes to sit at the four corners of a square, while identical frequencies appear across the diagonals.

We now let the adjacent modes with different frequencies $\omega$ and $\Omega$ undergo strong hopping interactions. These four interactions sitting at the four surrounding edges could be equal or different in magnitude, but must have precisely controlled phases as described later. We also let the pair of modes having identical frequencies to undergo weak parametric interactions. These constitute the diagonals of the square system, across which non-reciprocity or directional amplification is to be maintained. 

We observe that the requirement on strength of interactions, as being strong for hopping and weak for the parametric interactions are quite necessary for correct operation. For instance, in absence of either of the parametric interactions, the overall non-reciprocal feature of the system is lost, so that the presence of both of the diagonals is absolutely necessary for instrinsic nonreciprocity. However, phase of parametric interactions is irrelevant and may be dropped. 

Hence, the total number of interactions is six, comprising of four hopping and two parametric types. These complete the system configuration of diamond.

\begin{figure}
	\centering
	\includegraphics[width=3 in]{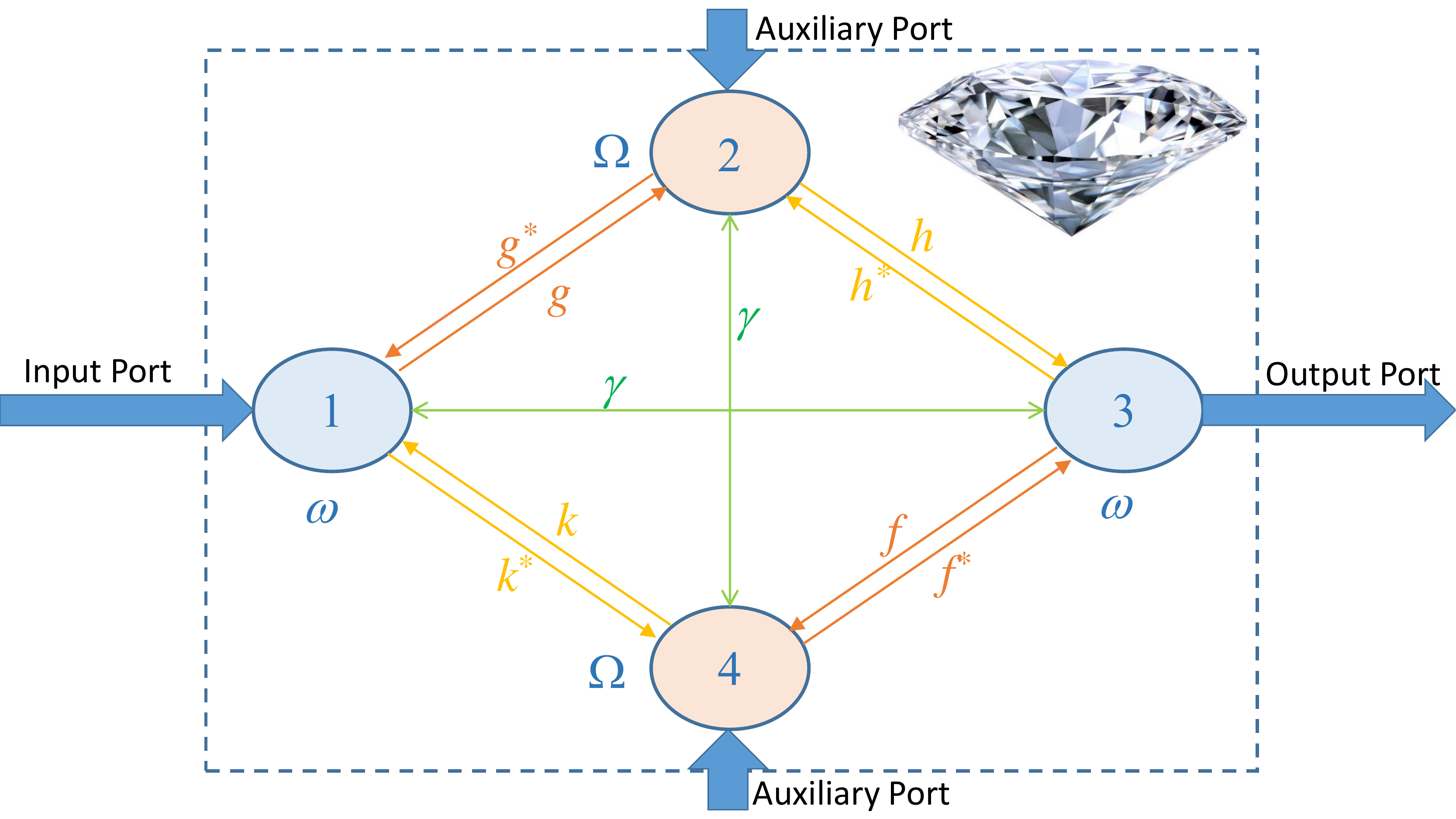}
	\caption{Schematics of the diamond-shaped non-reciprocal or directional amplifier device. Opposite ports across the diagonals at identical frequencies may be used as input/output pairs. \label{Fig1}}
\end{figure}

\subsection{Analysis}
The overall Hamiltonian $\mathbb{H}$ of the system consists of two non-interacting $\mathbb{H}_0$ and interacting parts $\mathbb{H}_i$, which in the regime of  Rotating-Wave-Approximation (RWA) may be described as 
\begin{eqnarray}\nonumber
\mathbb{H}_0&=& \hbar\omega(\hat{a}_1^\dagger\hat{a}_1+\hat{a}_3^\dagger\hat{a}_3+1)+\hbar\Omega(\hat{a}_2^\dagger\hat{a}_2+\hat{a}_4^\dagger\hat{a}_4+1),\\ \nonumber
\mathbb{H}_i&=&\hbar(g\hat{a}_1\hat{a}_2^\dagger+g^{*}\hat{a}_2\hat{a}_1^\dagger)+\hbar(f\hat{a}_3\hat{a}_4^\dagger+f^{*}\hat{a}_4\hat{a}_3^\dagger)\\ \nonumber
&+&\hbar(h\hat{a}_2\hat{a}_3^\dagger+h^{*}\hat{a}_3\hat{a}_2^\dagger)+\hbar(k\hat{a}_4\hat{a}_1^\dagger+k^{*}\hat{a}_1\hat{a}_4^\dagger)\\ \label{eq1}
&+&\hbar\gamma(\hat{a}_1\hat{a}_3+\hat{a}_1^\dagger\hat{a}_3^\dagger)+\hbar\gamma(\hat{a}_2\hat{a}_4+\hat{a}_2^\dagger\hat{a}_4^\dagger).
\end{eqnarray}
\noindent
Here, the bosonic operators satisfy $[\hat{a}_m,\hat{a}_n^\dagger]=\delta_{mn}$ and $[\hat{a}_m,\hat{a}_n]=0$, $g$, $h$, $f$, and $k$ are the strengths of hopping interactions across the edges, and also $\gamma$ is the strength of parametric interaction. The RWA holds if the pumps driving the hopping and parametric terms are resonant. For this one needs $\omega_{g}=\omega_{h}=\omega_{f}=\omega_{k}=|\omega-\Omega|$ and $\omega_\gamma=\omega+\Omega$. For this reason, $\omega\neq\Omega$ must hold if interactions are driven by electromagnetic drive fields. 

With the conditions stated in the above, the interaction (\ref{eq1}) requires using at least two coherent drive fields, which may be separated by splitters and reach the edges  at controlled phases or diagonals at arbitrary phases. As stated above, the phase of $\gamma$ has no effect on the system performance in any of the diagonal parametric interactions.

We may now use the input/output formalism \cite{18,19,20} to assign decay channels to each of the modes through the relevant ports. This will cause linewidths as $\Gamma_n$ with $n=1,2,3,4$ appear in the ultimate formulation, corresponding to the field Langevin equations. Because of parametric interactions, an $8\times8$ formalism needs to be used as 
\begin{equation}
\label{eq2}
\frac{d}{dt}\{a\}=[\textbf{M}]\{a\}-\sqrt{[\Gamma]}\{a_{\rm in}\},
\end{equation}
\noindent
where $\{a\}^{\rm T}=\{\hat{a}_1,\hat{a}_2,\hat{a}_3,\hat{a}_4,\hat{a}_1^\dagger,\hat{a}_2^\dagger,\hat{a}_3^\dagger,\hat{a}_4^\dagger\}$ is the system vector, and $[\Gamma]={\rm Diag}[\Gamma_1,\Gamma_2,\Gamma_3,\Gamma_4,\Gamma_1,\Gamma_2,\Gamma_3,\Gamma_4]$. Furthermore, $\{a_{\rm in}\}$ represents the input fields to the system at the ports. Similarly, one may define $\{a_{\rm out}\}$ as the output fields. These are related together as \cite{18,19,20,21}
\begin{equation}
\label{eq3}
\{a_{\rm out}\}=\{a_{\rm in}\}+\sqrt{[\Gamma]}\{a\}.
\end{equation}
\noindent
The scattering matrix formalism also requires that
\begin{equation}
\label{eq4}
\{a_{\rm out}\}=[\textbf{S}]\{a_{\rm in}\}.
\end{equation}
\noindent
By comparing (\ref{eq3}) and (\ref{eq4}) to (\ref{eq2}) we obtain the expression for the scattering matrix in terms of the system matrices as
\begin{equation}
\label{eq5}
[\textbf{S}(w)]=[\textbf{I}]-\sqrt{[\Gamma]}\left(iw[\textbf{I}]+[\textbf{M}]\right)^{-1}\sqrt{[\Gamma]},
\end{equation}
\noindent
where $w$ is the frequency of interest at either of the four ports and $[\textbf{I}]$ is the $8\times8$ identity matrix. 

The scattering matrix now $[\textbf{S}]$ is then well-defined if the system matrix $[\textbf{M}]$ is known. This can be obatined by using the Langevin equations given by
\begin{eqnarray}
\label{eq6}
\frac{d}{dt}\hat{a}&=&-\frac{i}{\hbar}[\hat{a},\mathbb{H}]\\ \nonumber
&-&[\hat{a},\hat{c}^\dagger]\left(\frac{1}{2}\Gamma\hat{c}+\sqrt{\Gamma}\hat{a}_{\rm in}\right)+\left(\frac{1}{2}\Gamma\hat{c}^\dagger+\sqrt{\Gamma}\hat{a}_{\rm in}^\dagger\right)[\hat{a},\hat{c}],
\end{eqnarray}
\noindent
where $\hat{c}$ is any system operator, which is here taken to be the same as $\hat{a}$ to comply with (\ref{eq3}). Now, once the scattering matrix $[{\bf S}(w)]$ is known from (\ref{eq5}), a measure of intrinsic non-reciprocity in transmission between ports 1 and 3 at the given frequency $w$, can be defined as
\begin{equation}
\label{eq7}
R(w)=\frac{1}{2}\left[\left|\frac{S_{31}(w)}{S_{13}(w)}\right|^2+\left|\frac{S_{13}(w)}{S_{31}(w)}\right|^2\right].
\end{equation}
All remains now, is to have the system matrix $[{\bf M}]$ known. This is found from the Hamiltonian (\ref{eq1}), Langevin equations (\ref{eq6}), and (\ref{eq2}) as
\begin{widetext}
\begin{equation}
\label{eq8}
[{\bf M}]=\begin{bmatrix}
	-i \omega-\frac{1}{2}\Gamma_1 & -i g^*& 0& -i k& 0& 0& -i \gamma & 0\\
	-i g& -i \Omega-\frac{1}{2}\Gamma_2 & -i h^*& 0& 0& 0& 0& -i \gamma \\
	0& -i h& -i \omega-\frac{1}{2}\Gamma_1 & -i f^*& -i \gamma & 0& 0& 0\\
	-i k^*& 0& -i f& -i \Omega-\frac{1}{2}\Gamma_2 & 0& -i \gamma & 0& 0\\
	0& 0& +i \gamma & 0& +i \omega-\frac{1}{2}\Gamma_1 & +i g& 0& +i k^*\\
	0& 0& 0& +i \gamma & +i g^*& +i \Omega-\frac{1}{2}\Gamma_2 & +i h& 0\\
	+i \gamma & 0& 0& 0& 0& +i h^*& +i \omega-\frac{1}{2}\Gamma_1 & +i f\\
	0& +i \gamma & 0& 0& +i k& 0& +i f^*& +i \Omega-\frac{1}{2}\Gamma_2 
\end{bmatrix}.
\end{equation}
\end{widetext}
\noindent
Here because of geometric considerations, we have assumed that $\Gamma_1=\Gamma_3$ and $\Gamma_2=\Gamma_4$ to further reduce the number of system parameters. We may later also assume $|f|=|g|$ for further simplification, but the optimum results are dependent on the application. Phases of $f$ and $g$ should be different to obtain the best results. Now, we are all set, and may proceed to the calculation of non-reciprocity function (\ref{eq7}) in terms of frequency $w$.
 
\section{Results}

We assume that $\omega=2\pi\times 1 {\rm GHz}$ and $\Omega=2\pi\times 2 {\rm GHz}$ represent the resonant frequencies of ports 1,3 and 2,4, respectively. Here, two cases can be distinguished, being referred to as the intrinsic versus extrinsic non-reciprocities. The intrinsic case is characterized by zero pump power into ports 2 and 4, which are at frequency $\Omega$. This is satisfied by putting $a_{{\rm in},2}=a_{{\rm in},4}=0$. The measure of non-reciprocity (\ref{eq7}) then applies. The extrinsic case is defined as the situation where there are input powers at ports 2 and 4 at frequency $\Omega$, which reads $a_{{\rm in},2}\neq 0$ and $a_{{\rm in},4}\neq 0$. The measure of non-reciprocity (\ref{eq7}) is senseless and should be redefined.

Choice of parameters are quite typical and accessible for microwave superconducting circuits. There are too many degrees of freedom in the proposed configuration to investigate them all, so in order to keep this number minimal, it has been tried to choose equal magnitudes for variables wherever possible.

\subsection{Intrinsic Non-reciprocity}

For the moment, we also notice that the strength of hopping interactions over all four edges should be different to disturb the equilibrium of bridge configuration, in such a way that either $|gh|\neq|fk|$ or $|gk|\neq|fh|$. This criteria, however, are neither necessary nor sufficient, but rather facilitate obtaining a strong non-reciprocity if both are satisfied. A simple choice which satisfies both is $|g|=|h|=|k|=2\pi\times 1{\rm MHz}$ and $|f|=2\pi\times 10{\rm MHz}$. Generally, the hopping interactions should significantly exceed parametric interactions, and here we choose $\gamma=2\pi\times 300{\rm kHz}$. Quality factors at ports 1,3 and 2,4 are respectively set equal to $Q_1=Q_3=2\times 10^3$ and $Q_2=Q_4=10^3$, corresponding to coupling rates $\Gamma_1=\Gamma_3=2\pi\times 500 {\rm kHz}$ at ports 1,3 and $\Gamma_2=\Gamma_4=2\pi\times 2 {\rm MHz}$ at ports 2,4. These values are quite typical and might be adjusted with ease to obtain the desirable characteristics. 
\begin{figure}[htp]
	\centering

	\subfloat{\includegraphics[width=3 in]{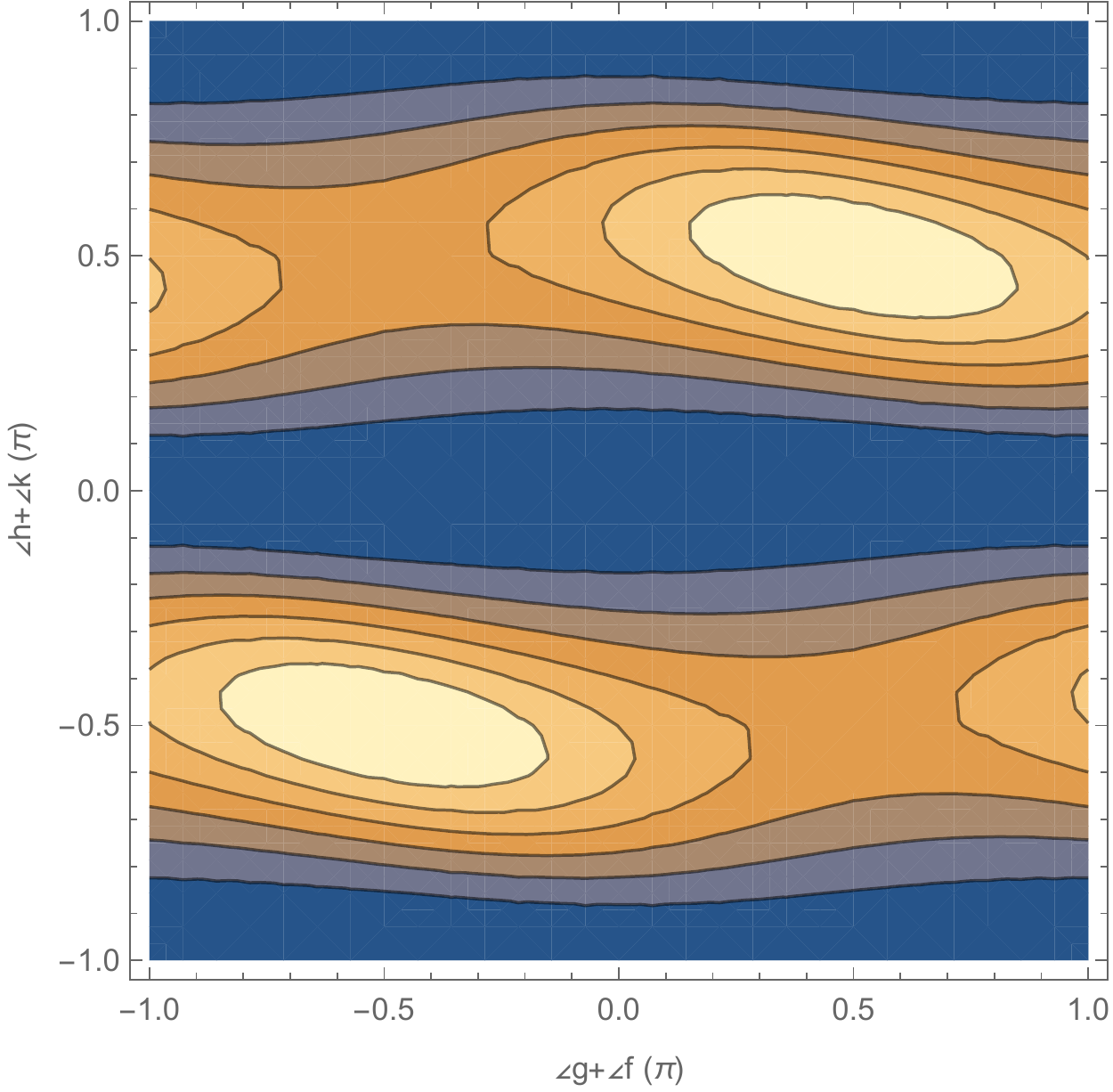}} 
	\subfloat{\includegraphics[width=0.5 in]{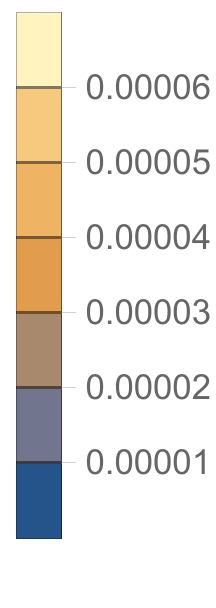}}

	\caption{The maximum non-reciprocal ratio is only a function of round-trip phase $\theta$. Extrema of non-reciprocity between ports 1 and 3 is obtained at $\theta=\pm\pi$ corresponding to $\angle g=\angle h=\angle f=\angle k=\pi/4$. \label{Fig2}}
\end{figure}
It is necessary to adjust the phases of hopping coupling rates $\angle g$, $\angle h$, $\angle f$, and $\angle k$. In principle, the sum of these four parameters should determine any non-reciprocal behavior, as it is related to the round-trip phase $\theta=\angle g+\angle h+\angle f+\angle k$. In Fig. \ref{Fig2}, the prominent effect of these phases on the non-reciprocity (\ref{eq7}) is illustrated. The response with respect to the change of sign of $\theta$ is unchanged. As it can be seen, the maxima of non-reciprocal ratio $R(w)$ at $w=\omega$ is obtained when $\theta=\pm\pi$, corresponding to the typical choice $\angle g=\angle h=\angle f=\angle k=\pi/4$. This confirms the requirement of destructive interference across the loop as the necessary condition for non-reciprocity. 

\begin{figure}[htp]
	\centering
	\includegraphics[width=3 in]{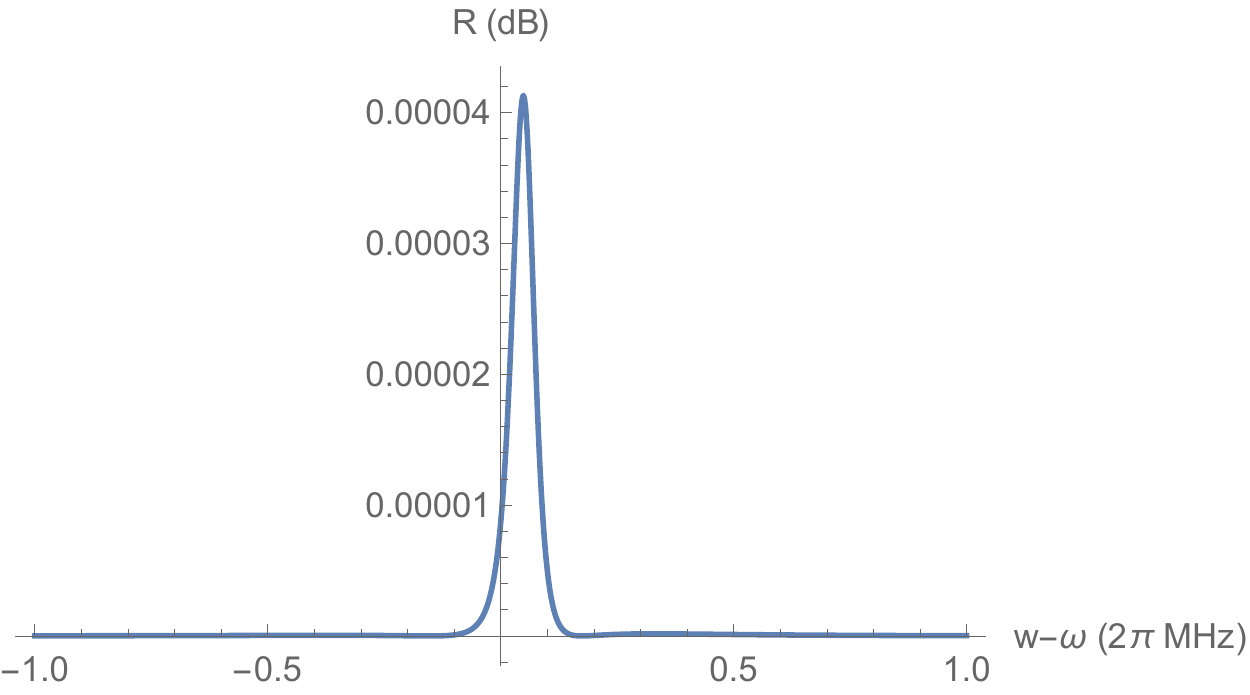}
	\caption{Unoptimized variation of non-reciprocity ratio (\ref{eq7}) across the resonant frequency $\omega$. The maximum value is rather small, but may be significantly enhanced by careful selection of parameters.\label{Fig3}}
\end{figure}
\begin{figure}[htp]
	\centering
	\includegraphics[width=3 in]{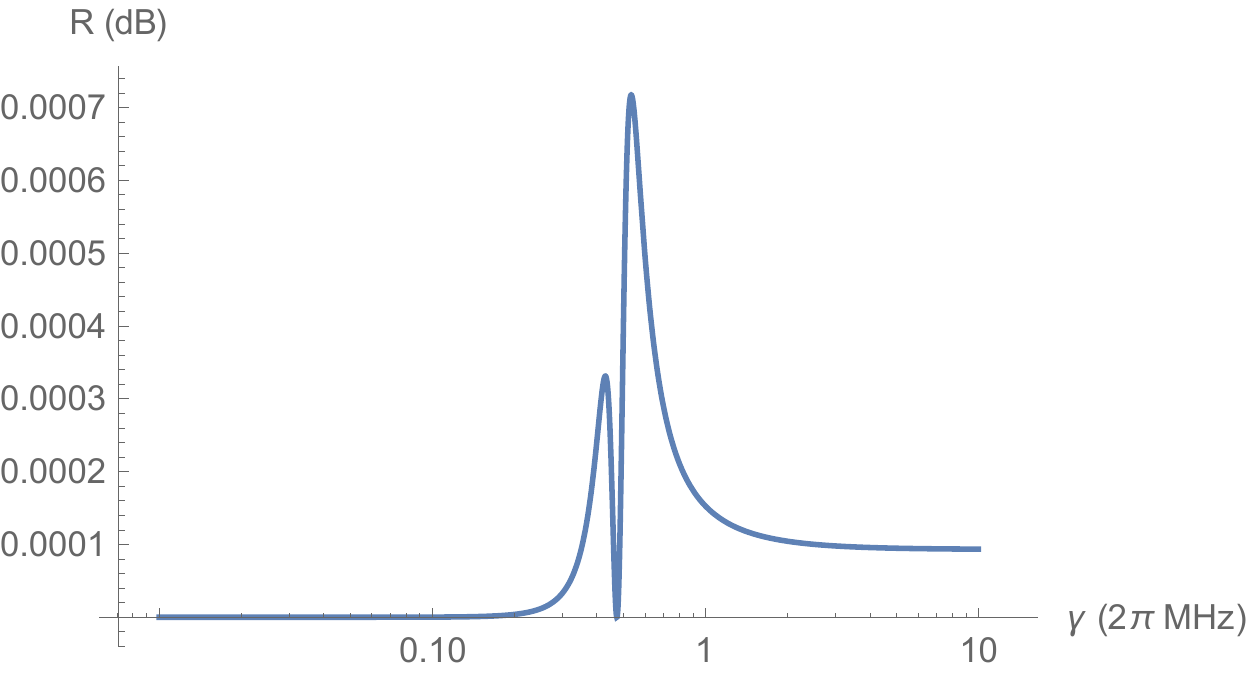}
	\caption{Unoptimized variation of non-reciprocity ratio (\ref{eq7}) versus strength of parametric interactions $\gamma$. This plot clearly shows that the choice of optimal $\gamma$ is limited to a narrow range of frequencies. \label{Fig3a}}
\end{figure}
\begin{figure}[htp]
	\centering
	\subfloat{\includegraphics[width=3 in]{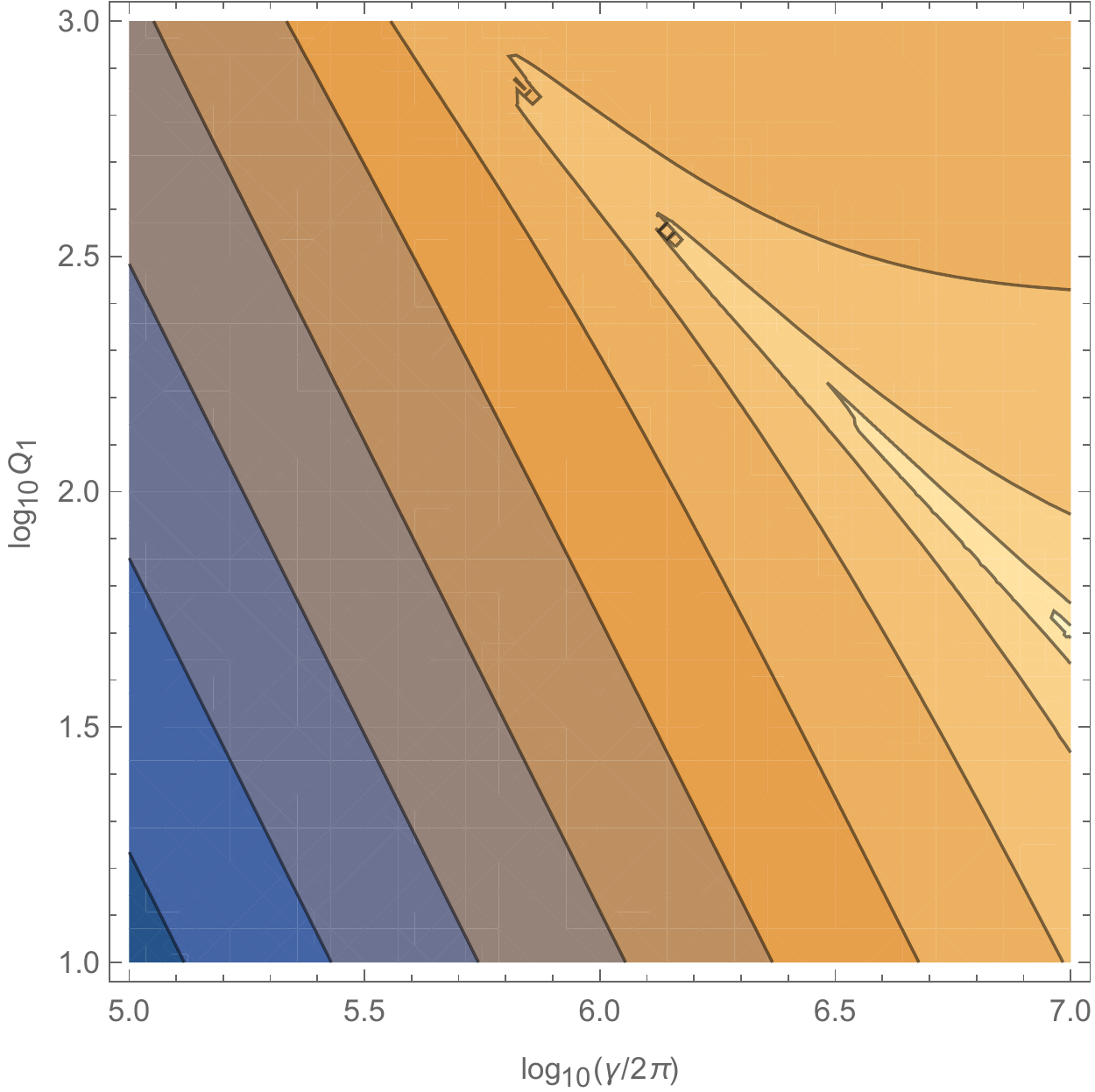}} 
	\subfloat{\includegraphics[width=0.5 in]{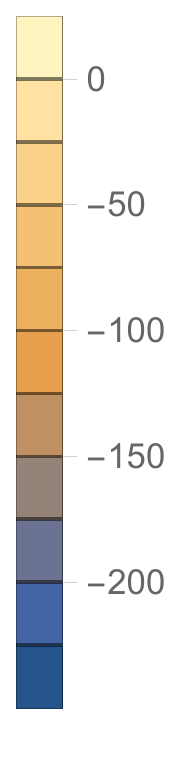}}
	
	\caption{Optimization of non-reciprocity ratio (\ref{eq7}) versus quality factor $Q_1=Q_3$ and the parametric interaction gain $\gamma$. The optimum value occurs at $Q_1=Q_3=51.286$ and $\gamma=2\pi\times 10{\rm MHz}$ at which $R(\omega)=3.652$. Choice of $Q_2=Q_4$ is much less relevant.\label{Fig4}}
\end{figure}
\begin{figure}[htp]
	\centering
	\includegraphics[width=3 in]{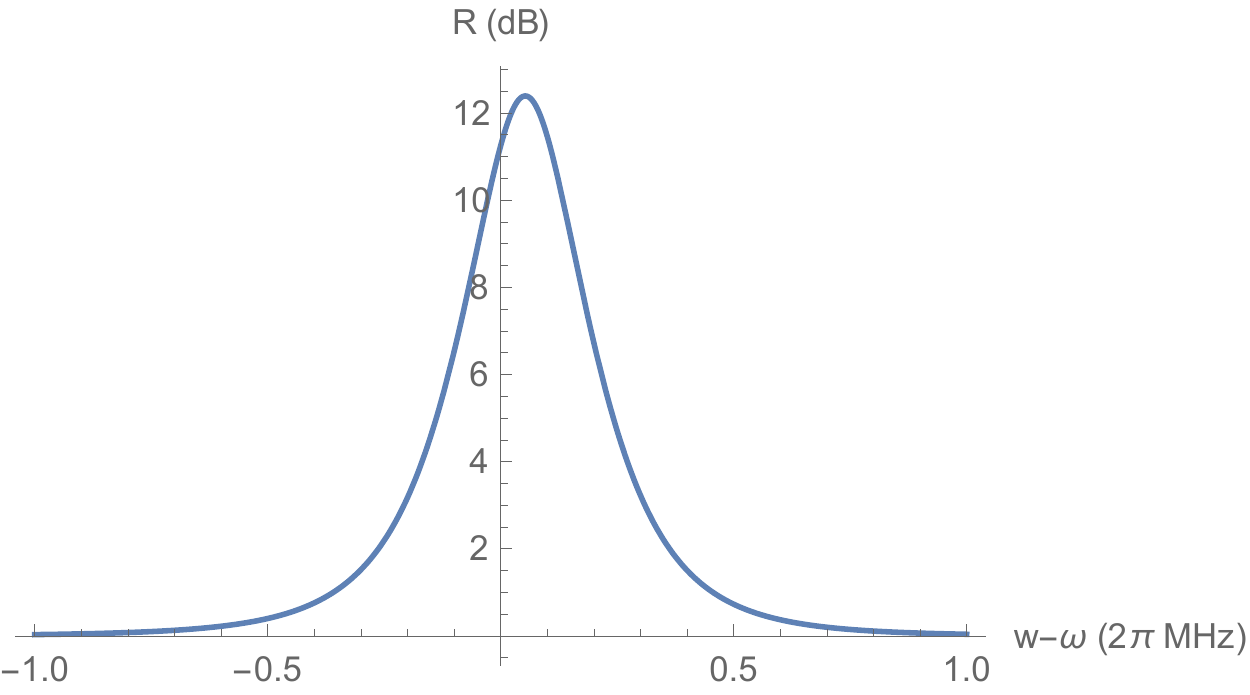}
	\caption{Optimized non-reciprocity ratio (\ref{eq7}) across the resonant frequency $\omega$. The maximum is blue detuned to the amount of $53{\rm kHz}$ and is equal to $R_{\rm max}=12.39{\rm dB}$.\label{Fig5}}
\end{figure}
\begin{figure}[htp]
	\centering
	\includegraphics[width=3 in]{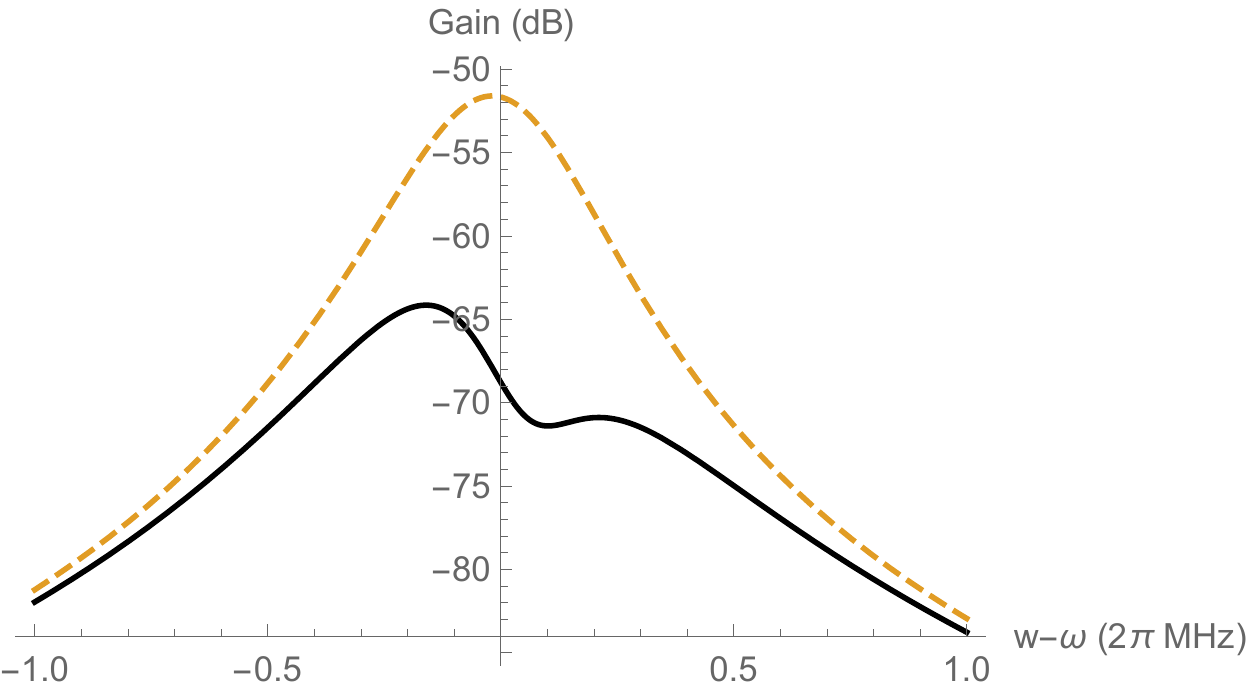}
	\caption{Optimized forward (dashed) and backward (solid) gains across the resonant frequency $\omega$.\label{Fig6}}
\end{figure}
\begin{figure}[htp]
	\centering
	\subfloat{\includegraphics[width=3 in]{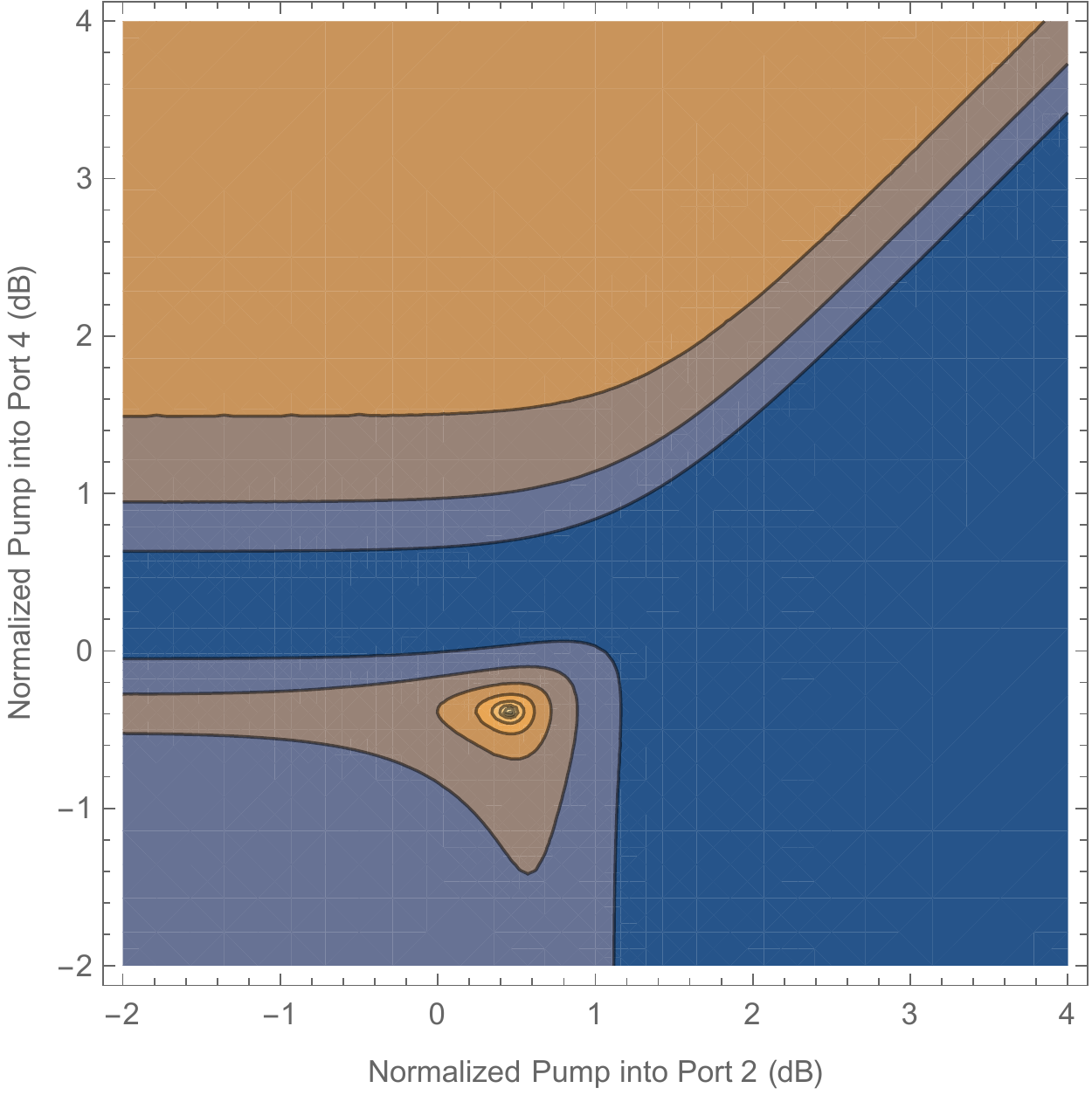}} 
	\subfloat{\includegraphics[width=0.5 in]{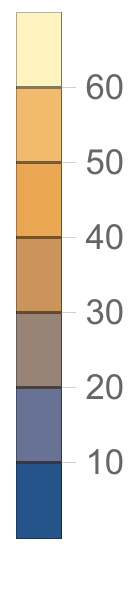}}
	\caption{Optimized input powers at ports 2 and 4 to obtain maximal non-reciprocity. Pumps are at frequency $\Omega$.\label{Fig7}}
\end{figure}
\begin{figure}[htp]
	\centering
	\includegraphics[width=3 in]{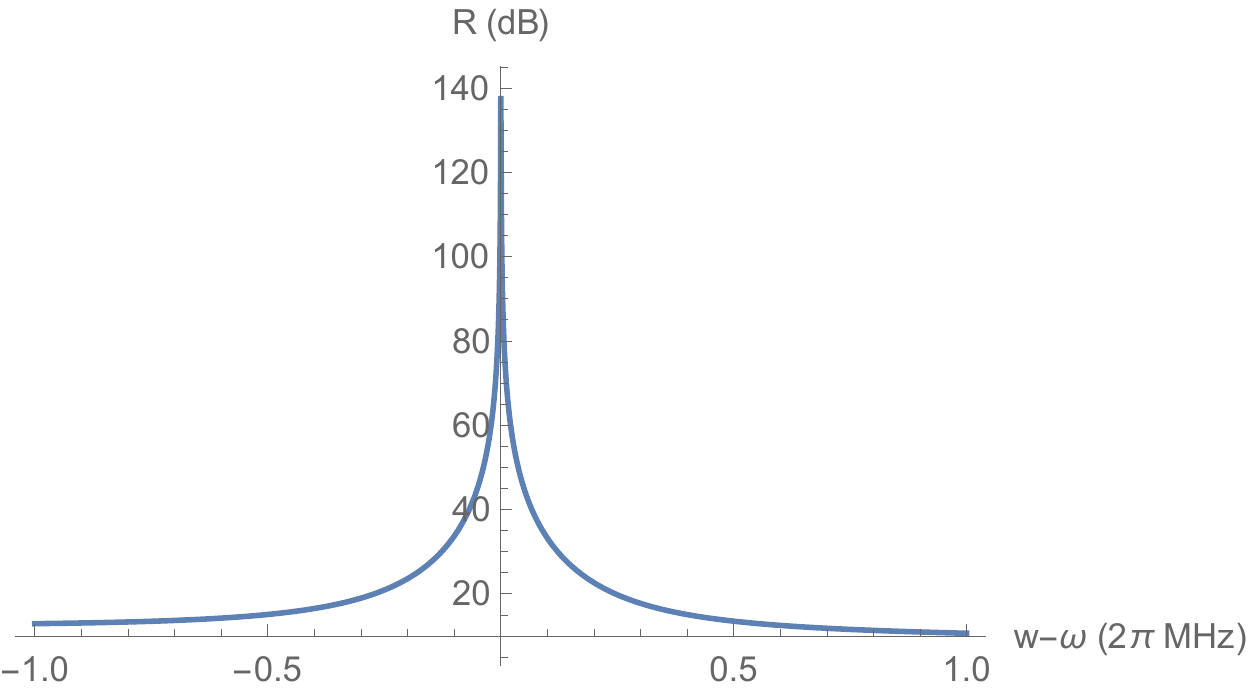}
	\caption{Optimized extrinsic non-reciprocity with pump powers fed into ports 2 and 4.\label{Fig8}}
\end{figure}
\begin{figure}[htp]
	\centering
	\includegraphics[width=3 in]{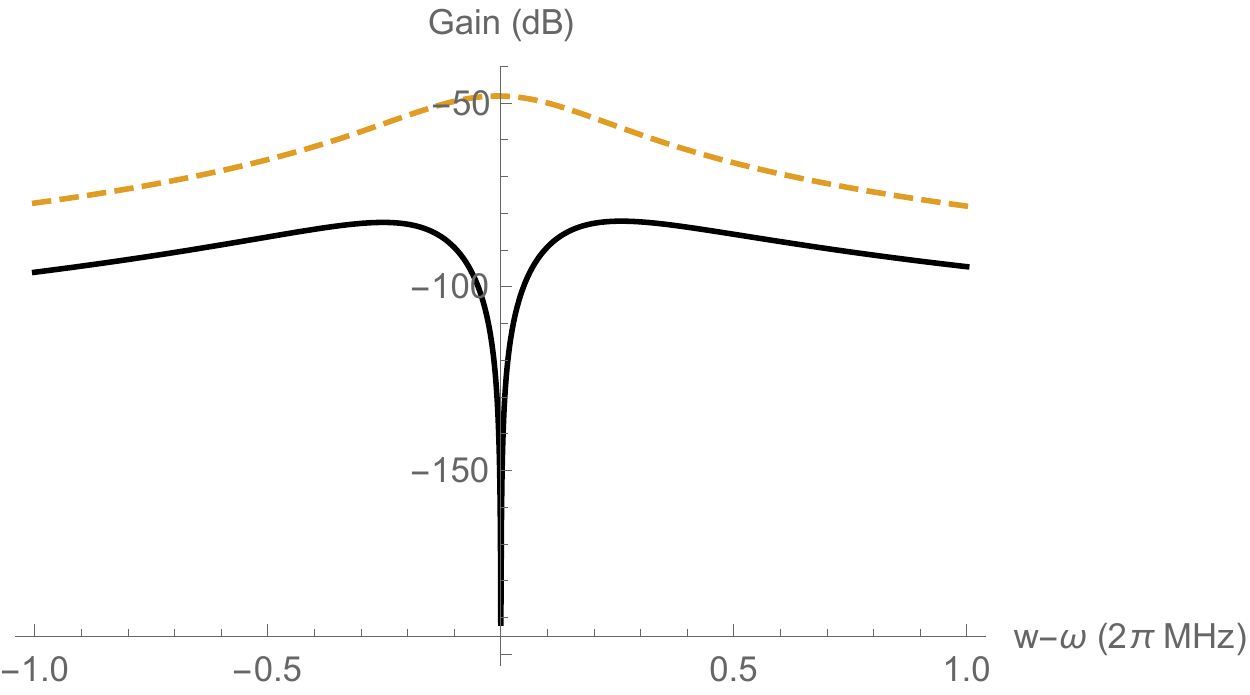}
	\caption{Optimized extrinsic forward (dashed) and backward (solid) gains across the resonant frequency $\omega$. The reverse gain has largely been suppressed with the aid of pumps at $w=\omega$.\label{Fig9}}
\end{figure}
\begin{figure}[htp]
	\centering
	\includegraphics[width=3 in]{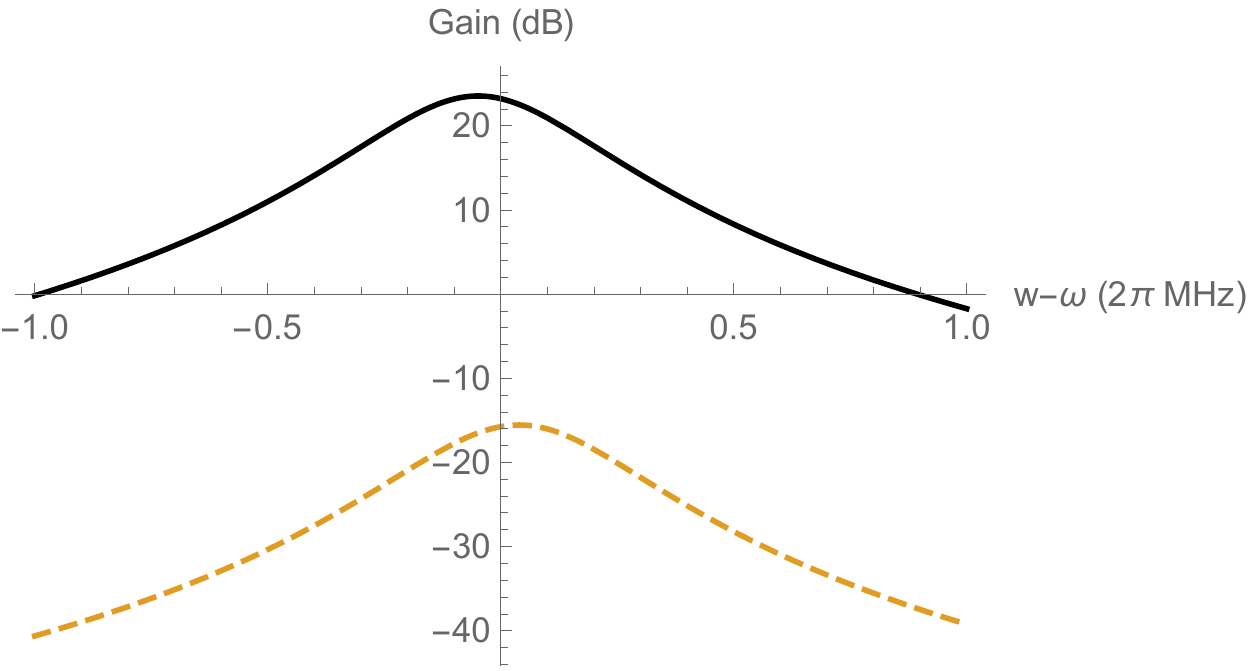}
	\caption{Directional Amplification extrinsic forward (dashed) and backward (solid) gains across the resonant frequency $\omega$. The $-3{\rm dB}$ gain window is roughly $\pm1{\rm MHz}$. \label{Fig10}}
\end{figure}

It has been observed that the hopping and parametric interactions are both needed to achieve intrinsic non-reciprocity. However, while the choice of hopping strength is much less sensitive, the choice of the parametric interaction strength is highly non-trivial. In order to show this, we plot and include a typical behavior as illustrated in Fig. \ref{Fig3a}. This plot reveals that the strength of parametric interaction is optimum in a very limit range of frequencies. A choice of lower or higher frequencies could totally destroy the non-reciprocal behavior.

In Fig. \ref{Fig4} the variation of $R(w)$ versus frequency $w$ around the resonance frequency $\omega$ of ports 1 and 3 is shown. The maximum is slightly blue-detuned but is very small. It is possible to optimize this value by appropriate selection of $Q_1=Q_3$ and $\gamma$. For this reason, we fix $Q_2=Q_4=10^4$ and make a contour plot of $R(w)$ in terms of $Q_1=Q_3$ and $\gamma$. This is shown in Fig. \ref{Fig4}. Optimal values are $Q_1=Q_3=51.286$ and $\gamma=2\pi\times 10{\rm MHz}$. This yields the maximum non-reciprocity of $R_{\rm max}=12.39{\rm dB}$, which is significantly enhanced comparing to the maximum value of only $4\times10^{-5}{\rm dB}$ in Fig. \ref{Fig4}. 

It should be stressed out that these optimized values have been obtained for the above assumed values of port frequencies $\omega$, $\Omega$ as well as hopping interaction rates $g$, $f$, $h$, and $k$. It is expected that by appropriate adjusting these values even much higher non-reciprocity could be obtained.

Notably, the intrinsic non-reciprocity is not useful for directional amplification, as the calculated forward and backward gains shown in Fig. \ref{Fig6} are much smaller than unity. This can be resolved by means of extrinsic configuration discussed in the following.

\subsection{Extrinsic Non-reciprocity}

In the extrinsic configuration, there is a non-zero pump power at frequency $\Omega$ into ports 2 and 4. The amplitudes of these could be optimized to obtain the maximal response. We here proceed with the last set of optimized values obtained for the intrinsic configuration to calculate the optimal pump powers. Normalized to input power incident to the port 1 $a_{{\rm in},1}$, the extrinsic measure of non-reciprocity could be now similarly redefined as
\begin{equation}
\label{eq9}
R=\frac{1}{2}\left(|W|^{+2}+|W|^{-2}\right),
\end{equation}
\noindent
in which 
\begin{equation}
\label{eq10}
W=\frac{S_{31}+S_{32}\bar{a}_2+S_{34}\bar{a}_4}{S_{13}+S_{12}\bar{a}_2+S_{14}\bar{a}_4},
\end{equation}
with $\bar{a}_2=a_{{\rm in},2}/a_{{\rm in},1}$ and $\bar{a}_4=a_{{\rm in},4}/a_{{\rm in},1}$ being the normalized auxiliary pumps into ports 2 and 4. 

By making a contour plot of $R(w)$ as shown in Fig. \ref{Fig7} we are able to obtain the optimal pump values into ports 2 and 4. These values are highly non-trivial, found as $\bar{a}_2=2.844$ and $\bar{a}_4=0.4121$.

The optimal extrinsic non-reciprocal response against detuning as been plotted in Fig. \ref{Fig8}. The peak non-reciprocal response has been increased very significantly to a very large amount in excess of 130dB.

\subsection{Directional Amplifier}

It is fairly easy to obtain directional amplification by manipulating the pump powers at ports 2 and 4. While the optimal values as shown in Fig. \ref{Fig7} can boost the non-reciprocal behavior by 120dB, the directional amplification needs a directional gain exceeding 0dB. This is contingent on non-optimal values of pump powers, in fact, and can be easily obtained. 

\begin{figure}[htp]
	\centering
	\includegraphics[width=3 in]{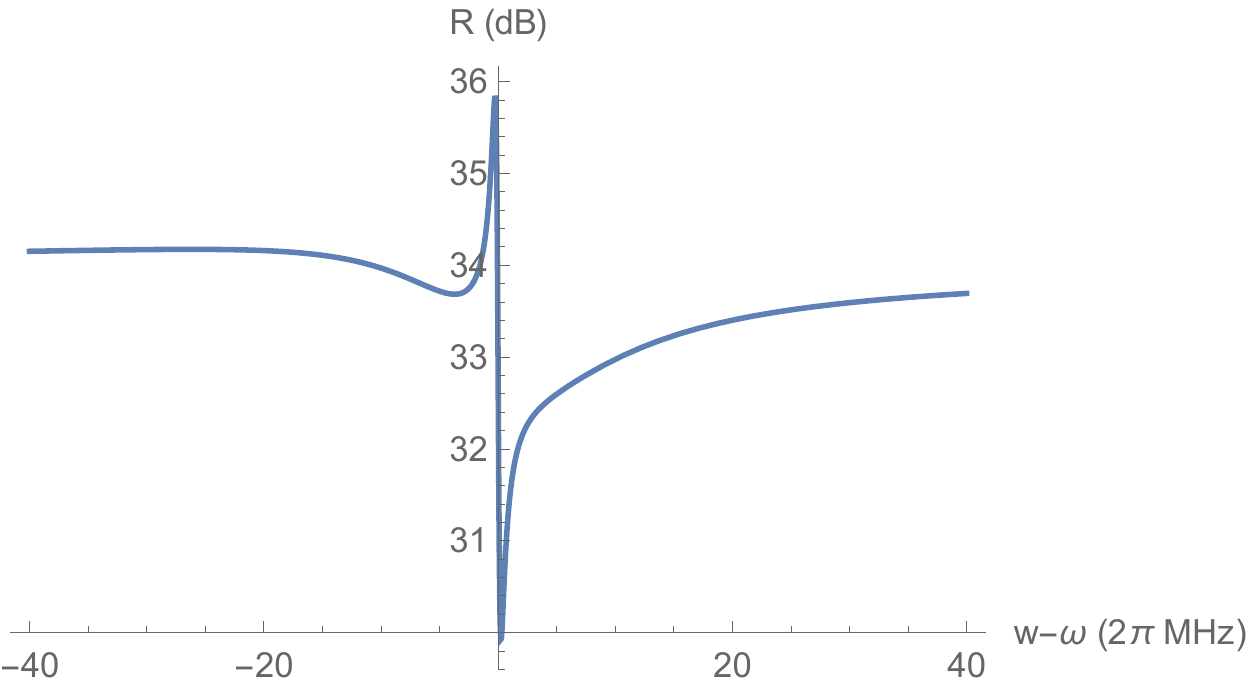}
	\caption{Extrinsic wide-band non-reciprocity. Blue- and red-detuned frequencies would observe respectively at least 30dB and 33.6dB of non-reciprocity. \label{Fig11}}
\end{figure}

Here, we may suppress the pump into port 2, while pumping hard into port 4 at frequency $\Omega$, say at $\bar{a}_4=10^2$. This will easily lead into the highly directional gain as depicted in Fig. \ref{Fig10}, with a wide-band non-reciprocity well exceeding 30dB as shown in Fig. \ref{Fig11}. The last plot can be extended over the frequency range by a factor of 10, and the non-reciprocity persists well.

The gain spectra shown in Fig. \ref{Fig10} almost scale with the pump power at port 4, leaving the non-reciprocity response in Fig. \ref{Fig11} unchanged. That implies by correctly staging two of such diamonds, and using the extremely high extinction in Fig. \ref{Fig9} and the directional gain in Fig. \ref{Fig10}, there is potential to achieve high directional amplification up to 50dB at $\bar{a}_4=10^4$ (or still more by increasing $\bar{a}_4$), while having a strongly non-reciprocal response easily as high as 170dB. However, this goal will need careful treatment of reflections at the conjoining port.

\section{Conclusions}

In summary, we have demonstrated the feasibility of non-reciprocal response in a diamond shaped system. In order to break the symmetry, it has been needed to have four modes all interacting with the rest. Each mode is coupled via hopping interaction to the nearest neighbors with different phases and magnitudes. However, non-neighboring modes need to undergo parametric interactions. These parametric interactions have been noticed to be quite necessary for non-reciprocal response and could not be removed, however, their relative phases are insignificant. Optimal interaction phases for obtaining maximum non-reciprocity ratio at the resonant frequency were obtained, amounting to a round-trip phase of $\pm\pi$. Quality factors of input/output ports versus parametric interaction rate could be significantly enhanced by optimization. Two basic intrinsic or unpumped, and extrinsic or pumped forms were discussed to achieve non-reciprocal behavior. With the aid of pumps, it was shown that the non-reciprocal behavior could be enhanced by around 120dB. For the directional amplification, an isolation as high as 40dB could be demonstrated with a forward/backward gains of $\pm20{\rm dB}$. Cascading two of such diamonds has the potential to offer large isolation and forward gains.

%\section*{Acknowledgement}
%The author wishes to thank Dr. Alexey Feofanov at \'{E}cole Polytechnique F\'{e}d\'{e}rale de Lausanne (EPFL) for discussions of the subject. The author is being supported in part by Research Deputy of Sharif University of Technology over a sabbatical visit. This research has been supported by the Laboratory of Photonic and Quantum Measurements (LPQM) at EPFL. The author is indebted to Dr. Kirstin Friedrich for overwhelming encouragement and enthusiasm regarding this work.

%\begin{IEEEbiographynophoto}{Sina Khorasani (S'98-M'05-SM'09)} received his M.Sc. and Ph.D. degrees from Sharif University of Technology respectively in 1996 and 2001, both in Electrical Engineering, where he is a Full Professor. He has been with School of Electrical and Computer Engineering at Georgia Institute of Technology as a Postdoctoral (2002-2004) and Research Fellow (2011-2012). He is now with \'{E}cole Polytechnique F\'{e}d\'{e}rale de Lausanne as a Visiting Professor. His active research areas include quantum optics and photonics, and quantum electronics. Dr. Khorasani is a Senior Member of IEEE.
%\end{IEEEbiographynophoto}
\end{document}